\documentclass[a4paper,11pt]{article}
\usepackage{jheppub}
\usepackage[usenames,dvipsnames]{xcolor}
\hypersetup{
    pdfencoding=unicode,
	colorlinks=true,
	urlcolor=Maroon,
	linkcolor=RoyalBlue,
	citecolor=OrangeRed,
	pdfdisplaydoctitle=true,
	pdfstartview=FitH,
	linktocpage=true
}
\usepackage{tikz}
\usetikzlibrary{shapes.misc}

\title{\boldmath Infrared limit of left-handed string at genus one}

\author[a]{Yuqi Li}
\affiliation[a]{C.~N.~Yang Institute for Theoretical Physics,\\ State University of New York, Stony Brook, NY 11794-3840}

\emailAdd{yuqi.li@stonybrook.edu}
\preprint{YITP-SB-2023-36}

\abstract{We extend the left-handed string formalism at one-loop level to focus on only the infrared limit, where the Green's function for the left-handed string is expanded around the cusp of the modular parameter. This expansion leads to the separating degeneration limit of a Riemann surface corresponding to a sphere and a torus connected by a long tube. The well-behaved short-distance behavior of the Green's function requires all marked points to be inserted on the sphere. Analogous to the tree-level calculations, we obtain Dirac $\delta$-functions by integrating out the anti-holomorphic variables. The constraints embedded in these $\delta$-functions, associated with the marked points on the sphere part of the Riemann surface, are the same Scattering Equations at the tree-level. After the integration over the modular parameter, we observe the expected pattern of the infrared divergence, consistent with the one-loop results from the box diagram calculations in field theory.}

\begin{document}
\maketitle
\flushbottom
\section{Introduction}
This paper is closely tied and directly connected to \cite{Li:2023oyq} in the study of the derivation of one-loop generalizations of the CHY amplitude from the left-handed string. In this work, we develop an alternative treatment, in which the infrared behavior of four-point the box amplitudes are derived directly, by taking the $\tau\rightarrow i\infty$ limit of the modular parameter to enable us to omit the zero modes of the Green's function. In 2015, Siegel \cite{Siegel:2015axg} introduced a novel method for calculating the tree-level scattering amplitude of massless particles in string theory, known as the left-handed string approach, which is founded on the chiral gauge transformation (Hohm-Siegel-Zwiebach (HSZ) gauge \cite{Hohm:2013jaa}) of the worldsheet coordinates of the closed string. It was immediately recognized that the left-handed approach is equivalent to the Cachazo-He-Yuan (CHY) formula \cite{Cachazo:2013hca,Cachazo:2013iea,Cachazo:2014nsa,Cachazo:2014xea} at tree-level. Subsequent detailed calculations have extended the left-handed string framework to the supersymmetric case \cite{Li:2017emw}. The CHY formula is also recognized as being derivable from conventional string amplitudes through the insertion of $\delta$-functions, as shown in various works \cite{Dolan:2013isa,Bjerrum-Bohr:2014qwa,Dolan:2014ega, Dolan:2015iln}. This was furthered by Mason and Skinner (MS) \cite{Mason:2013sva}, who embedded $\delta$-functions into each vertex operator within the ambitwistor string framework. Similarly, Berkovits \cite{Berkovits:2013xba} generalized the calculation of the scattering amplitude using pure spinor superstring formalism with $\delta$-functions inserted on vertex operators in the infinite-tension limit ($\alpha ' \rightarrow 0$). The discovery is also extended to tensile left-handed string at tree-level \cite{Azevedo:2019zbn,LipinskiJusinskas:2019cej,Guillen:2021mwp,Guillen:2021nky,Jusinskas:2021bdj}. 

At one-loop level, corresponding to a genus-one Riemann surface, the original CHY formula is not applicable. Some attempts to bridge this gap include explorations with the ambitwistor string at one-loop \cite{Adamo:2013tsa,Casali:2014hfa,Geyer:2015jch}. In \cite{He:2015yua}, the authors used the forward limit to obtain the one-loop Scattering Equations. Further discussions and developments regarding the one-loop integrand on the Riemann surface are available in \cite{Tourkine:2019ukp,Geyer:2021oox}. Similar to the technique in field theory to cut a loop into the tree graph, these attempts above are formulated in the non-separating degeneration limit \cite{Witten:2012ga,Witten:2019ylx,Witten:2019uux,Tourkine:2019ukp} of the Riemann surface. In this limit, the Riemann surface's genus decreases by one, while incorporating two new external massive states at specific marked points, characterized by loop momenta. Thus, the tree-level CHY formula, now including two extra massive external legs, can be employed to compute the one-loop amplitude with the integration over the loop momentum left to be performed.

In this work, we continue our explorations of the left-handed string approach at one-loop level by initiating our analysis from the infrared limit directly. Specifically, we will develop an expansion of the Green's function around the cusp of modular parameter $\tau$, in the $\tau\rightarrow i\infty$ limit. The remarkable feature of this asymptotic limit is to eliminate the need to include the zero modes. Thus, we will only need to consider the non-zero modes of the Green's function. This marks a departure from our previous work \cite{Li:2023oyq}, where we keep both zero and non-zero modes with the modification induced by HSZ gauge choice. Near the cusp, the limit $\tau_2\rightarrow\infty$ can be interpreted as a separating degeneration, deforming the Riemann surface into a configuration where a sphere and a torus are connected by a long tube, with all external states inserted on the sphere. Since all the external states are localized on the sphere, the Green's function's well-behaved short-distance behavior permits the reduction of the logarithm of the Jacobi theta function to a simple logarithm as the Green's function seen at tree-level. Following the HSZ prescription, the same tree-level CHY $\delta$-function emerges when the external states' coordinates are assigned as $(z_1,z_2,z_3,z_4)=(0,1,z,\infty_\tau)$, where $z$ satisfies the tree-level Scattering Equations, and $z_4\sim\infty$ is linked to the modular parameter $\tau=\tau_1+i\tau_2$ through additional constraints. Further calculations indicate that one feasible constraint of $z_4$ is an extra $\delta$-function of $z_4$. With the redefinition $\tau_2\sim t_1t_2$, the remaining integrals over $t_1$ and $t_2$ yield the expected infrared divergence of four-point scattering in the dimensional regularization. Finally, it is promising that this method can be straightforwardly generalized to higher points as well as to the higher-loop level, corresponding to the separating degeneration limit of a Riemann surface of higher genus.\\
\textbf{Note Added:} This paper is single-authored and separated from \cite{Li:2023oyq} due to the unfortunate circumstance that one author of \cite{Li:2023oyq} was unable to continue the work. I would like to acknowledge the special contributions of George Sterman and Martin Rocek to this work.

\section{HSZ at the infrared}
Recall the definition of Hohm-Siegel-Zwiebach (HSZ) gauge choice \cite{Siegel:2015axg}; the original HSZ gauge choice for the worldsheet coordinate transformation is:
\begin{eqnarray}\label{eq:def}
\chi:\,z\rightarrow \mathbf{z}=\sqrt{1+\beta}z,\,\, \,\, \bar z \rightarrow \mathbf{\bar z}=\frac{1}{\sqrt{1+\beta}}(\bar z -\beta z).
\end{eqnarray}
In the HSZ gauge, the Lagrangian canonically transforms into
\begin{eqnarray}
\mathcal{L}=-\frac{1}{2}[\beta(\bar\partial X)(\bar\partial X)+(\bar\partial X)(\partial X)].
\end{eqnarray}
As in the conventional string theory, the Green's function for the torus reads
\begin{eqnarray}\label{eq:Green}
 G(z_{ij}|\tau)=
 -\ln\left|\frac{\theta_1(z_{ij}|\tau)}{\theta'_1(0|\tau)}\right|^{2}+2\pi\frac{(\operatorname{Im}
   z_{ij})^2}{\operatorname{Im}\tau}\,,
\end{eqnarray}
where $z_{ij}=z_i-z_j$ and $\theta_1(z_{ij}|\tau)$s are ordinary Jacobi $\theta$ functions,
\begin{eqnarray}
\theta_1(z|\tau)=2iq^{\frac{1}{8}}\sin(\pi z)\prod^{\infty}_{j=1}(1-q^j)(1-e^{2\pi i z}q^j) (1-e^{-2\pi i z}q^j)
\end{eqnarray} 
with $\tau=\tau_1+i\tau_2$ the modular parameter of the torus parametrized $\tau_1=\mathrm{Rm}[\tau]$ and $\tau_2=\mathrm{Im}[\tau]$. And we also define $q=e^{2\pi i\tau}$ and $\theta'_1(z_{ij}|\tau)$ as the derivative of the Jacobi $\theta$ function. At the infrared limit, the expansion of the Green's function around the cusp of modular parameter $\tau$ corresponding to $\tau_2\rightarrow\infty$ only keeps the logarithmic part of the Green's function,
\begin{eqnarray}\label{eq:limit}
 G(z_{ij}|\tau)\overset{\tau_2\rightarrow\infty}{\longrightarrow}
 -\ln\left|\theta_1(z_{ij}|\tau)\right|^{2}-\ln\left|\theta'_1(0|\tau)\right|^{2}.
\end{eqnarray}
The $z$ independent $\theta'_1(0|\tau)$ term will be cancelled out due to the momentum conservation. Thus, we will omit this term in the following calculations. As in \cite{Li:2023oyq}, the tamed short-distance behavior of the Green's function allows us to perform the partial differential on the logarithm to obtain the same tree-level left-handed string Green's function, denoted by $G_\intercal(z_{ij}|\tau)$, namely,
\begin{eqnarray}\label{eq:flip}
G_{\intercal}(z|\tau)&\overset{\beta\gg1}{\Longrightarrow}&\frac{\bar z}{\beta}\frac{\partial}{\partial z}G_\intercal(z|\tau)=-\frac{\bar z}{\beta z}.
\end{eqnarray}
Comparing with \cite{Siegel:2015axg}, (\ref{eq:flip}) leads to the tree-level Scattering Equations as expected. Before delving deep into the detailed calculations, we would like to discuss the mathematical and physical consequences of the infrared limit.
As discussed in \cite{DHoker:2020prr}, any degeneration of the Riemann surface, both separating and non-separating, can be locally interpreted as pinching along the A-cycle. Each pinch will produce two more external states, correspondingly.

In the discussions of \cite{Casali:2014hfa,Geyer:2015jch,Geyer:2021oox}, the authors have used the non-separating degeneration limit to obtain the one-loop scattering amplitude. In this limit, since the Riemann surface is not separated, the genus of the surface decreases by one, with two extra external points punctured on the Riemann surface, namely,
\begin{eqnarray*}
\mathcal{M}_{g,n}\overset{\mathcal{D}_{\text{nonsep}}}{\Longrightarrow}\mathcal{M}_{g-1,n+2},
\end{eqnarray*}
with $\mathcal{M}_{g,n}$ the Riemann surface with genus $g$ and $n$ punctures. Here, the $\mathcal{D}_{\text{nonsep}}$ denotes the non-separating degeneration limit of the Riemann surface. In order to make use of the tree-level Scattering Equations, one will have to associate the two extra external states with some loop momenta and integrate out the loop momentum at the end of the calculation.

In the separating degeneration limit, the Riemann surface $\mathcal{M}_{g,n}$ separated in two surfaces of genus $g_1$ and $g_2$ marked by $n_1+1$ and $n_2+1$ punctures, respectively,
\begin{eqnarray*}
\mathcal{M}_{g,n}\overset{\mathcal{D}_{\text{sep}}}{\Longrightarrow}\mathcal{M}_{g_1,n_1+1}\times\mathcal{M}_{g_2,n_2+1},
\end{eqnarray*}
where $g_1+g_2=g$, $n_1+n_2=n$ and the symbol $\mathcal{D}_{\text{sep}}$ represents the separating degeneration limit of the Riemann surface. Here  the two additional external states are distributed onto the two surfaces, which are connected by a long tube parametrized by the modular parameter $\tau$. As long as the tube is long enough ($\tau_2\rightarrow\infty$), the two surfaces become increasingly distant and effectively decoupled. Our asymptotic limit of the Green's function, as defined in equation (\ref{eq:limit}), remains valid by placing all marked points on the sphere. And this approach is consistent with the HSZ left-handed condition, which require the well-behaved short-distance behavior while $z\rightarrow0$ in (\ref{eq:flip}). For genus one Riemann surface with $n$ punctures, we have
\begin{eqnarray}
\mathcal{M}_{1,n}\overset{\mathcal{D}_{\text{sep}}}{\Longrightarrow}\mathcal{M}_{1,1}\times\mathcal{M}_{0,n+1}.
\end{eqnarray}
In contrast to the scenarios in the non-separating degeneration cases discussed earlier, the introduction of loop momentum is unnecessary. 

Let's take four-point amplitude as an example. And the rest of the paper will focus on the four-point amplitude. To apply the four-point CHY formula on the sphere, it is necessary to identify those extra points on both surfaces:
\begin{itemize}

	\item On the sphere side, we can asymptotically set the configuration of four points as $(z_1,z_2,z_3,z_4)=(0,1,z,\infty_\tau)$, with $z=-\frac{s_{13}}{s_{12}}$\footnote{The Mandelstam variables are defined as $s_{ij}=k_i\cdot k_j$.} satisfying the tree-level Scattering Equations. The notation $\infty_\tau$ requires an additional constraint on $z_4$. Detailed calculations suggest that one suitable constraint is to add an extra $\delta(\frac{\sqrt{3}}{2}t_1z_4-\tau_2)$, where $t_1$ is a finite real parameter. When $\tau_2\rightarrow\infty$, the support in $\delta$-function localizes $z_4=\frac{2}{\sqrt{3}}\tau_2/t_1:=\infty_\tau$, and the volume of the conformal Killing Group ($\mathrm{Vol(CKG)}$) is modified by an integration over the extra $\delta$-function, namely, $\mathrm{Vol(CKG)}=\int dz_4\delta(\frac{\sqrt{3}}{2}t_1z_4-\tau_2)|z_{12}z_{24}z_{41}|^2$.
	\item On the torus side, we define $t_2:=\frac{2}{\sqrt{3}}\tau_2/t_1$ and specify the external point such that it reparametrizes the modular parameter $\tau$ as $\tau=t_1-\frac{1}{2}+\frac{\sqrt{3}}{2}it_1t_2$, where $\tau_1=t_1-\frac{1}{2}$ and $\tau_2=\frac{\sqrt{3}}{2}t_1t_2$. This approach involves interpreting the modular parameter $\tau$ as a function of $t_1$ and $t_2$ by reparametrization, thereby fixing the modular parameter and swapping the roles of the modular parameters $(\tau_1,\tau_2)$ and the external point $(t_1,t_2)$. Consequently, the integration domain for $t_1$ and $t_2$ is restricted to $t_1\in[0,1]$ and $t_2\in[1,\infty]$, respectively.
\end{itemize}
After performing the integration over $t_1$ and $t_2$, we obtain the anticipated infrared divergence, which will be detailed below.

\section{Four-point Amplitude}
As in \cite{Li:2023oyq}, we continue to use the heterotic string in our calculations. The vertex operators are defined as in \cite{Schlotterer:2016cxa, Mafra:2022wml}:
\begin{eqnarray}
V^{a_i}(z_i,\bar z_i)&=&V_{\textrm{SUSY}}(z_i,\epsilon_i,k_i)\bar J^{a_i}(\bar{z}_i)e^{ik_i\cdot X(z_i,\bar z_i)}\nonumber\\
V^{\text {g }}(z_i, \bar{z}_i) &:=& -\frac{1}{2 \alpha^{\prime}} \tilde\epsilon_i^\mu \bar\partial X_\mu(\bar z_i) V_{\mathrm{SUSY}}(z_i) e^{ik_i \cdot X(z_i, \bar z_i)}.
\end{eqnarray}
Here, $V_{\textrm{SUSY}}(z_i,\epsilon_i,k_i)$ is the supersymmetric gauge multiplet associated with the $i^{\text {th }}$ external particle with polarization $\epsilon_i$, and momentum $k_i$, $\bar{J}^{a_i}(\bar{z}_i)$ represents the right-moving Kac-Moody current, and $\tilde{\epsilon}_i^\mu \bar{\partial} X\mu(\bar{z}_i)$ corresponds to the gravity counterpart of the multiplet, subject to transversality $(\tilde\epsilon \cdot k)=0$. The vector indices $\mu=0,1, \ldots, d-1$ span $d$ spacetime dimensions and the color $a_i$ refers to the $\mathit{SU}(N)$ generator $t^{a_i}$. In our analysis, the integrated and unintegrated vertex operators are not distinguished, and only the symbolic form of the correlation functions of the vertex operators is considered.\footnote{For more details on the supermultiplets and vertex operators, we refer the reader to \cite{Mafra:2022wml} and \cite{Gerken:2018jrq}.}

In the conventional string calculation, the four-point (super-)Yang-Mills amplitude, corresponding to the correlation function of four gauge vertex operators, is given by\footnote{In the conventional heterotic string, the measure part includes a factor of Dedekind eta function $\eta(\tau)$ \cite{Gerken:2018jrq}, which is omitted here because it is cancelled in the left-handed string \cite{Siegel:2015axg}.}
\begin{eqnarray}
A_4^{(1)}&=&\int d\tau_1 d\tau_2 \prod_{i=2}^4d^2z_i\frac{1}{\tau_2^{d/2-1}}\langle V^{a_1}(z_1,\bar z_1)V^{a_2}(z_2,\bar z_2)V^{a_3}(z_3,\bar z_3)V^{a_4}(z_4,\bar z_4)\rangle\nonumber\\
&=&\int\frac{d\tau_1 d\tau_2}{\tau_2^{d/2-1}} \prod_{i=2}^4d^2z_i \mathcal{A}_{KN}\langle \mathcal{K}(z_i)\rangle
\end{eqnarray}
after fixing one point by translation invariance, say $z_1=0$. In the second line, we define 
\begin{eqnarray}
\langle \mathcal{K}(z_i)\rangle&=&\langle\prod_{i=1}^4V_\text{SUSY}(z_{i},\epsilon_i,k_i)\rangle\langle \prod_{i=1}^4J^a(\bar z_{i})\rangle
\end{eqnarray}
and extract the Koba-Nielsen factor 
\begin{eqnarray}
	\mathcal{A}_{KN}=\exp\left[-\frac{1}{2}\sum_{i\neq j}\alpha'k_i\cdot k_jG(z_{ij}|\tau)\right].
\end{eqnarray}
In the separating degeneration of the left-handed string, we will impose the transformation corresponding to the HSZ gauge choice on the worldsheet coordinates. Thus, with the help of the deformed Green's function (\ref{eq:flip}), the Koba-Nielsen (KN) factor will be changed into
\begin{eqnarray}
	\tilde{\mathcal{A}}_{KN}=\prod_{i=1}^{4}\exp\left[-\frac{\bar z_i}{\beta}\sum_{j\neq i}\frac{\alpha'k_i\cdot k_j}{z_{ij}}\right].
\end{eqnarray}
Recall that \cite{Siegel:2015axg}
\begin{eqnarray}\label{eq:limit}
	\frac{1}{z}\rightarrow\frac{1}{\sqrt{\beta}}\frac{1}{z}, \,\,\,\,\,
	\frac{1}{\bar z}\rightarrow-\frac{1}{\sqrt{\beta}}\frac{1}{z}(1+\frac{1}{\beta}\frac{\bar z}{z}).
\end{eqnarray}
The short-distance behavior with $z_{ij}=z_i-z_j\sim0$ will force the anti-chiral part of the operator product expansion (OPE) to only depend on holomorphic variables in the leading order of $\frac{1}{\sqrt{\beta}}$. Since $\beta$ is a redundant gauge parameter to cancel out in the final result of the amplitude \cite{Siegel:2015axg,Li:2017emw}, we will omit it in the following calculations. 

As all the $z_i$ are localized on the sphere side of the separating degeneration limit, the integration over the $z_i$ and $\bar z_i$ will be performed on the Riemann sphere, with the modified volume $\mathrm{Vol(SL(2))}=dz^2_1d^2_2dz_4^2/\mathrm{Vol(CKG)}$ modded out. Let's use the configuration mentioned before, $(z_1,z_2,z_3,z_4)=(0,1,z_3,z_4)$, and link $z_4=\infty_\tau$ such that $t_2$ is always sufficiently large to ensure the validity of use of the solution to the tree-level Scattering Euqations. Putting all together, the amplitude reads
\begin{eqnarray}
A_4^{(1)}=\frac{1}{(\frac{\sqrt{3}}{2})^{d/2-2}}\int\frac{dt_1 t_1dt_2}{(t_1t_2)^{d/2-1}}\int\frac{\prod_{i=2}^4d^2z_i}{\mathrm{Vol(SL(2))}} \tilde{\mathcal{A}}_{KN}\langle \mathcal{K}(z_i)\rangle.
\end{eqnarray}
Denote $A_4^{\text{tree}}=\int\frac{\prod_{i=2}^4d^2z_i}{\mathrm{Vol(SL(2))}} \tilde{\mathcal{A}}_{KN}\langle \mathcal{K}(z_i)\rangle$ as the tree-level-like amplitude correpsponding to the sphere side of the separating degeneration limit, and we have
\begin{eqnarray}\label{eq:tree}
&&A^{\text{tree}}_4=\int dz_4\delta(\frac{\sqrt{3}}{2}t_1z_4-\tau_2)\mathrm{Vol(CKG)}\int dz_3d\bar z_3\exp\left[-\bar z_3\sum_{j\neq 3}\frac{\alpha'k_3\cdot k_j}{z_{3}-z_j}|_{z_4\rightarrow\infty_\tau}\right]\langle \mathcal{K}(z_i)\rangle\nonumber\\
&=&\frac{2}{\sqrt{3}t_1}(z_{12}^2z_{24}^2z_{41}^2)\int dz_3 \delta\left[-\alpha'(\frac{s_{31}}{z_{3}}+\frac{s_{32}}{z_3-1})\right]\frac{\textrm{tr}(t^{a_1}t^{a_2}t^{a_3}t^{a_4})}{z_{13}z_{32}z_{24}z_{41}}\left[\alpha's_{14}\right] \frac{A^{YM}(1,2,3,4)}{z_{12}z_{23}z_{34}z_{41}}|_{z_4=t_2}\nonumber\\
&=&-\frac{2}{\sqrt{3}t_1}\frac{\textrm{tr}(t^{a_1}t^{a_2}t^{a_3}t^{a_4})A^{YM}(1,2,3,4)}{\mathrm{Jac}(z_3)}\frac{s_{14}}{z_{13}z_{23}^2}\frac{z_{42}}{z_{43}}|_{z_4=t_2,z_3=z}\nonumber\\
&=&\left[-\frac{2}{\sqrt{3}t_1}\frac{t_2-1}{t_2-z}\right]\textrm{tr}(t^{a_1}t^{a_2}t^{a_3}t^{a_4})A^{YM}(1,2,3,4).
\end{eqnarray}
Here, $\textrm{tr}(t^{a_1}t^{a_2}t^{a_3}t^{a_4})$ comes from the OPE of Kac-Moody current, $\mathrm{Jac}(z_3)=\frac{s_{12}^3}{s_{13}s_{14}}$ is the Jacobian corresponding to the $\delta$-function, and $A^{YM}(1,2,3,4)$ is the color-ordered (super-)Yang-Mills amplitude. As in \cite{Li:2023oyq}, we already used the Kawai-Lewellen-Tye (KLT) relation to obtain the color-ordered amplitude \cite{Mafra:2010jq,Broedel:2013tta,Mafra:2010jq} in the second line. As already discussed in \cite{Li:2022tbz}, the left-handed string is relaxed from the constraint of critical dimension. Thus, at $d=4-2\epsilon$ dimension, keep only the leading divergence by ignoring the constant term in the numerator of (\ref{eq:tree}), and we have, up to a constant factor of powers of $\frac{\sqrt{3}}{2}$,
\begin{eqnarray}\label{eq:dimful}
A_4^{(1)}&=&-A^{\text{tree}}_4\int_0^1\frac{dt_1}{(t_1)^{d/2-1}}\int_1^\infty\frac{dt_2}{(t_2)^{d/2-2}}\frac{1}{t_2-z}\nonumber\\
&=&-A_4^{\text{tree}}\frac{1}{\epsilon}\lim_{b\rightarrow1}\int_{1}^{\infty}dt_2t_2^\epsilon\frac{1}{t_2-z}\frac{1}{(t_2-1)^{b-1}}\nonumber\\
&=&-A_4^{\text{tree}}\frac{1}{\epsilon}\frac{\Gamma(-\epsilon)\Gamma(1)}{\Gamma(1-\epsilon)}\, _2F_1(1,-\epsilon;1-\epsilon;z)
\end{eqnarray}
Here, $\, _2F_1(1,-\epsilon;1-\epsilon;z)$ is the Hypergeometric functions. In the sedond line, we have used the simlar calculation as in \cite{Smirnov:2006ry} to regularize the integral over $t_2$.\footnote{Strictly speaking, the integral domain of $t_2\in[1,\infty]$ contains infinitely many identical copies of the original integrals over $\tau_2$. However, this results in an overall factor that is independent of the dimensional regulator $\epsilon$. Thus, we can safely ignore this factor in our analysis.}
At $d=4-2\epsilon$ dimensions, our result (\ref{eq:dimful}) is not dimensionless. As seen in \cite{Bern:1987tw,Bern:1989fu,Bern:1990cu} and espescially in detailed in \cite{Bern:1991aq}, the coupling constant $g$ aquires a factor of $(\mu^2)^{\epsilon}$, where $\mu$ is the renormalization scale. This procedure leaves $\alpha'$ cancellation in (\ref{eq:tree}) not changed, but the $s_{14}$ term obtained from the KLT factor is modified by a correction of $(-s_{14})^{-\epsilon}$. The divergence can also come from the end point $1$ in the $t_1$ integral. After summing up the two terms,
$$\int_{0}^{1}dt_1\left[t_1^{\epsilon-1}+(1-t_1)^{\epsilon-1}\right]\approx\int_{0}^{1}dt_1t_1^{\epsilon-1}(1-t_1)^{\epsilon-1}=B(\epsilon,\epsilon)=\frac{\Gamma(\epsilon)^2}{\Gamma(2\epsilon)}$$ 
when $\epsilon\ll1$.
Define 
$$s=s_{12}=s_{34},\,\,\,\,\,\,t=s_{23}=s_{14},\,\,\,\,\,\,u=s_{13}=s_{24},$$
and sum up all the $s-t$ exchanged terms, we have the final crossing symmetric form of the amplitude,
\begin{eqnarray}
A_4^{(1)}&=&-A_4^{\text{tree}}\frac{\Gamma^2(\epsilon)\Gamma(-\epsilon)}{\Gamma(2\epsilon)\Gamma(1-\epsilon)}\nonumber\\
&\times&\left[(-t)^{-\epsilon}\, _2F_1(1,-\epsilon;1-\epsilon;1+\frac{t}{s})+(-s)^{-\epsilon}\, _2F_1(1,-\epsilon;1-\epsilon;1+\frac{s}{t})\right].
\end{eqnarray}
Comparing to \cite{Smirnov:2006ry}, there is a minus sign difference for all the $\epsilon$ in the $\Gamma$-functions and our results include one overall factor of $\Gamma(1-\epsilon)$ in the denominator, which will give rise to an exact $-\frac{1}{\epsilon}$ factor. Apart from these differences, the leading singular behavior of the Laurent expansion of $\Gamma$-functions implies that our result is consistent with the one-loop box diagram calculations in the field theory \cite{Bern:1991aq,Bern:1992em,Bern:1993kr}.

\section{Conclusion and Outlook}
In this paper, we further extend the one-loop left-handed string calculation in the infrared limit and the simplified Green's function suggests a potential separation of the surfaces. Indeed, the infrared limit, characterized by $\tau_2\rightarrow\infty$, can be interpreted as a separating degeneration limit that connects a sphere and a torus via a long tube.
By relocating all punctures to the sphere, we can employ the tree-level Scattering Equations, thereby reducing the number of integrals to just two: those over the reparametrized modular parameters $t_1$ and $t_2$. For the CHY-type Yang-Mills amplitude at the one-loop level, the Laurent expansion of dimensional regulator at $4-2\epsilon$ dimension of our result yields a pattern similar to that observed in field-theoretic calculations. It is crucial to emphasize that, in contrast to the ambitwistor calculations on the separating degeneration of the Riemann surface, our result is achieved without the need of introducing loop momentum.

Furthermore, based on this fundamental characteristic of the separating degeneration limit, we are prompted to explore the potential for extending our methodology to higher-loop scenarios. These scenarios would correspond to Riemann surfaces with higher genus, potentially allowing for the Bern-Dixon-Smirnov (BDS) \cite{Bern:2005iz} ansatz to be reformulated in a manner that relates to the genus of the Riemann surface:
$$M_4^{(g)}(\epsilon)=A_4^{(g)}/A^{\text{tree}}_4$$
with $g$ the genus of the corresponding Riemann surface at $g$-loop level. Thus, the leading singularity could be generated by
$\mathcal{S}^{(2g)}\approx \frac{1}{\epsilon^{2g}}+\ldots$. We would like to see the two-loop result as one example. 

\acknowledgments
Y.L. would like to express his heartfelt gratitude to George Sterman and Martin Rocek for their invaluable encouragement and support. This entire project could not have been accomplished without their contributions. Additionally, Y.L. appreciates the constructive conversations with Yao Ma, Lorenzo Magnea, Piotr Tourkine, Xiaojun Yao, and Peng Zhao. The research of Y.L. received the support from the National Science Foundation under Grant No. PHY-2215093.


\bibliographystyle{JHEP}
\bibliography{ref}
\end{document}